\documentclass[preprint]{vgtc}               

\graphicspath{{figures/}{pictures/}{images/}{./}} 
\vgtcinsertpkg
\usepackage{orcidlink}
\usepackage{times}                     

\usepackage{tabu}                      
\usepackage{booktabs}                  
\usepackage{lipsum}                    
\usepackage{mwe}                       

\newcommand{\revised}[1]{#1} 

\usepackage{mathptmx}                  

\onlineid{1002}

\vgtccategory{Research}

\title{Textarium: Entangling Annotation, Abstraction and Argument}

\author{Philipp Proff \orcidlink{0009-0009-1138-9027} \thanks{e-mail: philipp.proff@tu-berlin.de}\\ %
        \scriptsize Technical University Berlin \\ \scriptsize Berlin University of the Arts
        \and Marian Dörk \orcidlink{0000-0002-3469-7841} \thanks{e-mail: marian.doerk@fh-potsdam.de}\\
        \scriptsize University of Applied Sciences Potsdam} 

\teaser{
  \centering
  \vspace{-.5in}
  \includegraphics[width=\linewidth]{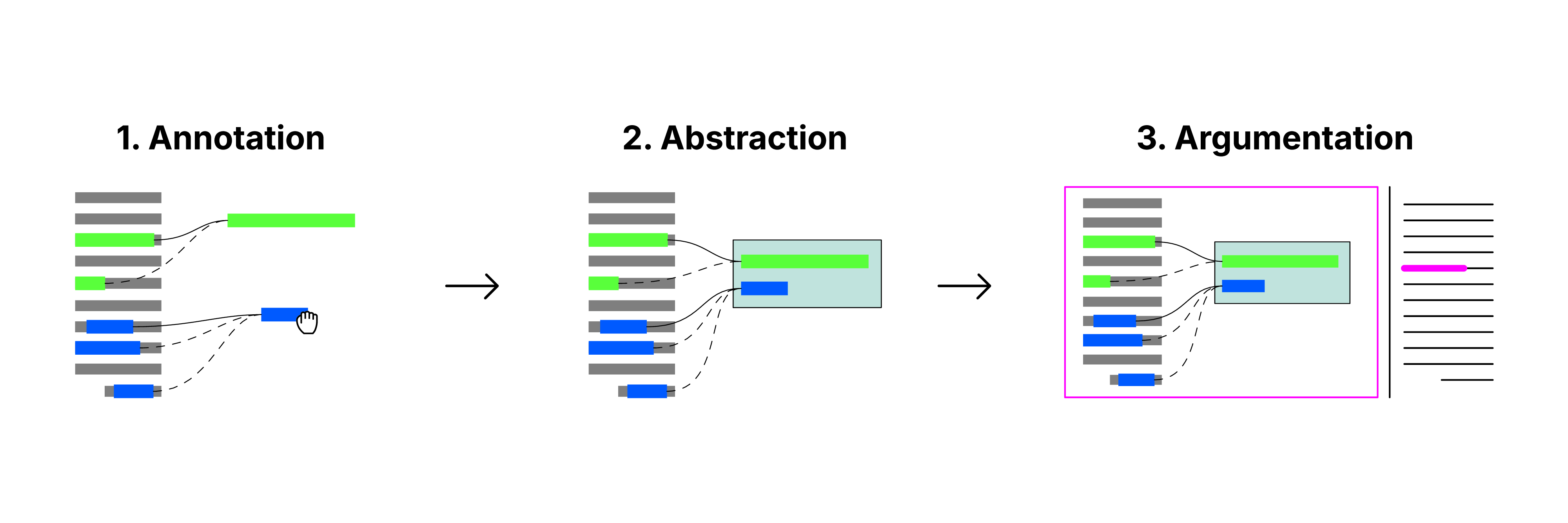}\vspace{-.5in}
  \caption{Workflow in the Textarium environment. User-curated annotations and abstractions get algorithmically augmented and can be embedded into an interactive article.}
  \label{fig:teaser}
}

\abstract{
We present a web-based environment that connects annotation, abstraction, and argumentation during the interpretation of text. As a visual interface for scholarly reading and writing, Textarium combines human analysis with lightweight computational processing to bridge close and distant reading practices. Readers can highlight text, group keywords into concepts, and embed these observations as anchors in essays. The interface renders these interpretive actions as parameterized visualization states. Through a speculative design process of co-creative and iterative prototyping, we developed a reading-writing approach that makes interpretive processes transparent and shareable within digital narratives.

} 


\begin{document}


\maketitle
\section{Introduction}
The process of text interpretation has changed many times over the course of history. With the emergence of the computer, the analysis and interpretation of text have been fundamentally transformed and computer-assisted text analysis has become an integral part of interpretative practices \cite{rockwell_hermeneutica_2017}. As natural language processing (NLP) and machine learning (ML) become better at processing textual data, research projects within the digital humanities (DH) explore their possible use-cases during the text interpretation process. The survey by Jänicke et al.~\cite{janicke_close_2015} has shown a division between top-down approaches extracting overall patterns from large of corpora~\cite{michel_quantitative_2011}, and bottom-up techniques that augment established reading techniques \cite{kehoe_emargin_2013}. However, there is a growing critique that distant reading methodologies, when used in isolation, can yield meaningless results \cite{da_computational_2019} and obscure ambiguous evidence~\cite{dimara_critical_2022}. 

An increasing number of projects is interested in combining close and distant reading techniques because the algorithmic analysis of complex relationships can benefit from human contextualization~\cite{hayles_how_2010}. Visualizations play a central role in this debate, as they can streamline and augment various analysis tasks \cite{liu_bridging_2019}. 
Data visualizations have the potential to bridge the gap between close and distant reading, facilitating a more holistic and critical interpretation process \cite{tversky_visualizing_2011}. DH scholar Johanna Drucker has even argued for acknowledging interactive visualizations as the central mode of knowledge production \cite{drucker_visualization_2020}.
There has also been growing interest in integrating interactive visualizations within other dissemination formats, to enhance their communicative qualities. Scrollytellings are a popular format allowing authors to create narrated stories that punctually use data visualization to depict concepts or provide visual evidence \cite{seyser_scrollytelling_2018}. 
The sciences have adopted interactive notebook environments, such as Jupyter Notebook or Observable, to integrate computational data analysis with research documentation and dissemination. In contrast, similar DH tools either specialize in text analysis or result presentation. This creates an opportunity for the design and development of tools that combine these interdependent modes of inquiry. Such tools could enable readers to interactively explore others’ text interpretation workflows and even start alternative explorations from their checkpoints. To address these aspirations we developed Textarium, an experimental web-based environment that integrates interpretation and authoring of text.

\section{Background}
Our work relates to previous research on interactive text visualizations, especially those combining close and distant reading, open-ended text visualizations, and interactive data stories.%

\textit{Text Analysis}. Novel text structures and formats, such as hypertext, have prompted new text analysis methodologies to address contemporary questions in literature. Hybrid forms of text analysis are needed, that can adapt to novel, nonlinear, digital text formats and build on the recent advancements in NLP. For Lev Manovich “computers and software are not just ‘technology’ but rather the new medium in which we can think and imagine
differently” \cite[p. 13]{manovich_software_2013}. Katherine Hayles exemplifies this idea by combining of close reading with social network platforms, such as Facebook, for text analysis \cite{hayles_how_2010}. Close reading is often split into an “inspectional” reading to get an overview of the text structure and an “analytical” reading for an in-depth analysis of characters, style, and inter-textual relations. Highlighting and annotation are important techniques in this process because they enable readers to actively engage with the text. Additionally, annotation allows the inscription of one’s own perspective into the analysis process and visually curate new knowledge~\cite{kalir_annotation_2021}. Most text analysis tools for close reading provide split-screen interfaces for the creation and modeling of hyperlinked annotations on a canvas \cite{craig_tashman_liquidtext_2011}. Many tools also allow scholars to visually alter the source text, to emphasize specific structures or word occurrences \cite{janicke_close_2015}.
Tools such as CATMA, developed in the context of literary studies, show how flexible annotation environments can accommodate a range of analytical approaches from exploratory to taxonomy-driven~\cite{gius_catma_2025}.
Unlike physical paper, digital tools allow the extraction, re-arrangement, and linking of text sections \cite{coles_empowering_2014}.
Due to the rapid advances in NLP and ML research, machines are becoming better at extracting meaning from texts, which is often referred to as machine reading \cite{eisenstein_introduction_2019}.
NLP algorithms are increasingly used in the DH to normalize historical morphological changes in words \cite{pettersson_normalisation_2013}, analyze word reuse patterns \cite{citron_patterns_2015}, and conduct comparative analyses for works by the same author \cite{ryan_comparative_2023}. The employed algorithms build upon the advances in recent decades from computational linguistics and information theory. They are built to be deterministic and produce consistent results. However, as natural language is a dynamic system, contemporary ML architectures are more flexible and produce probabilistic models which learn from large amounts of “wild” text data. Although, this allows the models to be more adaptable, they are also prone to produce inconsistent results and reproduce biases from their training data \cite{bolukbasi_man_2016}.

\textit{Text Visualization}. The visualization of textual data can support interpretative processes by transforming linguistic features into visual representations. 
Interactivity plays a central role in most text visualization techniques as it allows to reconfigure views and reorganize the contents as a form of direct interpretation \cite{armaselu_metaphors_2017}. A survey by Jänicke et al.~\cite{janicke_close_2015} outlines a range of visualization approaches that support both close and distant reading, highlighting the importance of visual annotation as a bridge between them.
Approaches in which the readers’ annotations serve as a starting point can be called bottom-up. Research in this area has explored “human-in-the-loop” techniques that allow readers to steer ML algorithms via annotations~\cite{koch_varifocalreader_2014} or support comparative analysis by providing split views of annotations from different analysis iterations \cite{wheeles_juxta_2013}. Conversely, tasks in which the machine initially analyzes the entire text can be called top-down. Top-down approaches typically utilize algorithms for semantic clustering and semantic mapping \cite{risch_text_2008}. 
For example, Voyant utilizes pre-defined computational analyses---such as word frequency, collocation, and topic modeling---to extract patterns from individual documents and large corpora~\cite{sinclair_voyant_2016}.
The results of the analysis can be made visible through a close reading view by adding annotations in the source text or creating interactive network visualizations for the exploration of relationships among textual entities \cite{paranyushkin_infranodus_2019}. Furthermore, environments like Stereoscope~\cite{kleymann2021towards} and TextBI \cite{masson_textbi_2024} feature multi-faceted visualizations that aim to make interpretive processes transparent and support the construction of scholarly arguments through visual means. Although, these approaches already support a wide-range of analysis tasks, most tools do not support the communication of findings within the visual analysis process. 

\textit{Interactive Data Stories}. Integrating interactive data visualizations into a narrative structure strikes a balance between a linear storytelling and open-ended exploration \cite{segel_narrative_2010}. One form of interactive data stories, scrollytellings allow authors to create a linear narrative and embed visualizations at specific points in the text \cite{seyser_scrollytelling_2018}. However, the implementation process remains labor-intensive, as most journalists lack training in data visualization and web development. Therefore, recent research has explored the possible improvements for making the format more attainable \cite{hohman_communicating_2020}. Several projects have attempted to streamline the creation process by integrating visualization creation and story authoring within a single environment \cite{conlen_idyll_2018}. 
Early experiments explored how networked annotation structures could support both interpretation and communication, pointing toward more integrated environments for scholarly reasoning~\cite{kirsch_tango_2018}.
An important aspect of such approaches is the parametrization of the interactive visualizations, which can be embedded into distinct URLs~\cite{eingartner_inflecting_2025}. This approach allows authors to integrate the reference and share specific states within Markdown-based writing tools, which are technically quite simple to use.

\section{Towards Visually Entangled Interpretation} 
With the increasing computational power of consumer-level computers, text analysis algorithms have become common in scholarly research. These algorithms can rapidly process, aggregate, and visualize large volumes of textual data. However, they continue to struggle with interpreting ambiguity, rhetorical nuance, and stylistic variation---aspects that are central to humanistic interpretation. In contrast, human readers, trained in close reading, apply various techniques for interpreting these rhetorical and stylistic devices. 

To support this interpretative process, scholars frequently use annotations to highlight text passages of interest, create abstractions of observed patterns, and add further contextual commentary. Visualization techniques can augment this practice by introducing lightweight “visual provocations” that direct attention to potentially meaningful areas of the text~\cite{esposito_reading_2022}. Yet, as computational models grow more complex, the visual outputs they generate become increasingly opaque. This makes it harder to understand the underlying processes, which is highly detrimental to trustworthy research.

We advocate for lightweight text analysis---such as pattern matching, stemming, or morphological similarity---that remains legible and controllable by scholars. These techniques can enrich the interpretive process without introducing excessive abstraction or algorithmic opacity. The visual form of annotations still allows readers to express conceptual relations and hierarchies, through graphical arrangements and elements, such as lines and boxes~\cite{tversky_visualizing_2011}. These augmentations support not only annotation, but also abstraction: the grouping and modeling of interpretative insights.
Both annotations and abstractions should remain anchored in the source text, and the interpretative acts that produced them should be reliably parameterized. These parameterized states can then be referenced, shared, and embedded in subsequent stages of scholarly work, including the construction of arguments and narratives.

Despite the recent research on exploratory text analysis tools and interactive storytelling, we know of no system offering an integrated environment that spans annotation, abstraction, and argument. In light of this, our work is guided by the following research question: 
\textit{How can a digital environment enable visual annotation and abstraction during text interpretation, while embedding the resulting arguments in interactive scholarly articles?}

\begin{figure*}[hbt!]
    \centering
    \includegraphics[width=\textwidth]{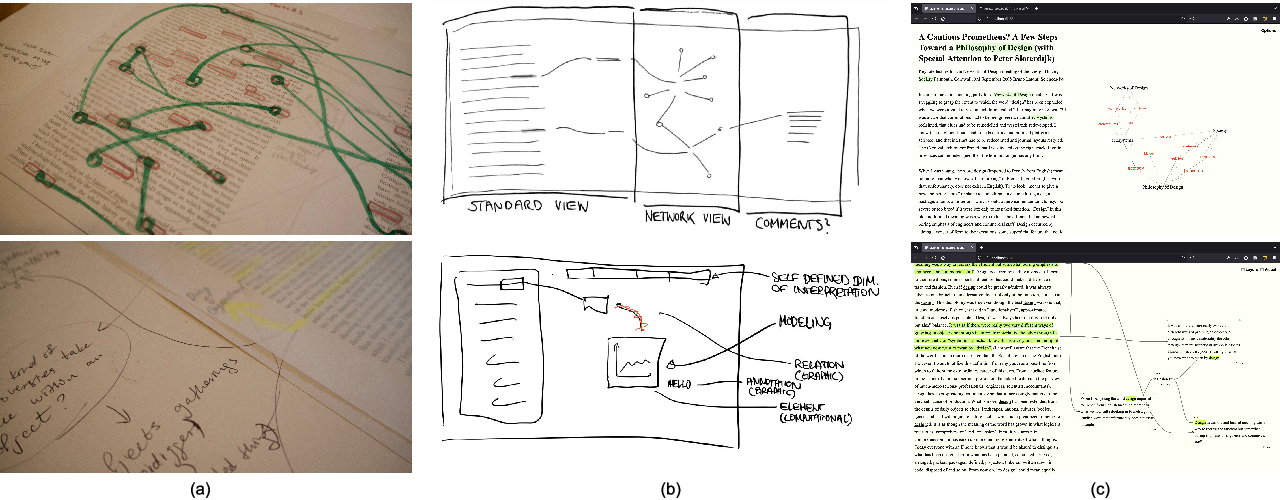}
    \vspace{-0.25in}
    \caption{Outcomes of the different research methods: (a) visualizations of annotations and abstraction of the co-design workshop participants, (b) sketches of possible interface layouts and functionalities and (c) prototypical implementations of functionalities.}
    \vspace{-0.2in}
    \label{fig:method-outcomes}
\end{figure*}
\subsection{Methodology}
Our research follows an interdisciplinary approach situated between digital humanities and data visualization. To support this, we combine three complementary methods: (1)~a co-design workshop, (2)~speculative sketching, and (3)~rapid prototyping---each incorporating critical reflection. The resulting design--critique cycles provided the evidential grounding for the final design decisions.

\textit{Co-Design Workshop}. Although this work is concerned with a digital environment, paper-based workshops are an efficient method for producing artifacts that prompt discussion and imagination \cite{bansemir_experience_2014}. In particular, co-design workshops can be fruitful for generating such artifacts, by bringing together an interdisciplinary group of participants~\cite{dork_co-designing_2020}. Our workshop was conducted with seven students from various academic backgrounds, including design, philosophy, architecture, cognitive science, and engineering. This diversity was intended to encourage and accommodate a broad range of analysis and visualization strategies and provide an interdisciplinary overview. All participants were asked to read and interpret a two-page excerpt from Bruno Latour \cite{latourCautiousPrometheusFew2008} within a time limit of 60 minutes. “Interpretation” was left undefined to encourage participants to fall back on their own strategies and associations. To document this process, all participants were asked to visually articulate their interpretation process on a DIN A3 poster by using materials and tools such as markers, pens, glue, string, transparent paper, and scissors. During the task, it was observed that most participants used markers to highlight words and sections of interest. Lines were among the most common graphical elements for indicating relationships (see \autoref{fig:method-outcomes}a). In the follow-up discussion, participants were asked to describe and interpret each others’ posters. This raised questions about the use of visual conventions for expressing sequentiality---such as left-to-right (in Western cultures) or top-down layouts. Some graphic elements followed a shared symbolic language, e.g., lines, while others were more arbitrary, e.g., colors, and therefore harder to decode. Visualizations were seen as helpful in forming analogies and metaphors to convey abstract concepts. Some participants pointed out the non-explicit associations for individuals which transformed visualizations into external cognitive anchors, yet potentially inaccessible to others. It was debated to what extent the visualizations should be self-explanatory. All participants agreed on the importance of “textual grounding” to maintain a connection between the visualizations and the respective sections in the source text. Lastly, participants noted an increased engagement with the text due to the associative connections formed between different visual analysis outcomes. Overall four key aspects were extracted from this discourse:

\begin{itemize}
    \item  Lines have a broadly shared foundation of meaning; colors and abbreviations are more ambiguous and harder to decode.
    \item Annotations were used to visually remix analysis outcomes with existing knowledge.
    \item Visual encodings of sequentiality, such as opacity or direction, are helpful for reading guidance.
    \item Visual references to the source text can enhance clarity, readability, and engagement.
\end{itemize}

\textit{Speculative Sketches}. The findings of the co-design workshop, along with prior research, prompted us to explore specific functionalities and visualizations that could be utilized to support the text interpretation process. As the design process produced many associations and ideas, sketching helped to visually develop potential interface elements. The sketching acted as a recursive process, where each sketch triggered new associations and ideas. While the explored text visualization techniques differed between the sketches, one interface layout stood out, as it facilitated a visually comparative view: the source text on the left and a space for abstraction on the right and visually connected through the center by line elements (see \autoref{fig:method-outcomes}b). 
Sketches that provided promising configurations of  functionalities were further explored by rapid prototyping to experiment with their potential and surface possible complications.

\textit{Rapid Prototyping}. 
Prototyping was used to test the viability of design iterations and uncover issues through interaction. Hinrichs et al. \cite{hinrichs_defense_2019} describe early-stage visualization prototypes as “sandcastles”---quick to create, aesthetically provocative, and useful  for reflecting on design decisions. For this phase, we selected \revised{an excerpt from} Bruno Latour’s “A Cautious Prometheus?”~\cite{latourCautiousPrometheusFew2008} as the source text to test analysis and visualization functionalities. One major exploration utilized word embedding models \cite{mikolov_efficient_2013,bojanowski_enriching_2016} to generate relational networks between annotations and possibly related words in the source text (see \autoref{fig:method-outcomes}c, upper image). Although, this produced thought-provoking associations, the embeddings demonstrated notable biases, which led to confusing relations and inconsistent results. We also tested various approaches to create abstractions from annotations. A mixed-initiative approach proved promising for identifying recurring themes, which were visually grouped to suggest higher-level concepts (see \autoref{fig:method-outcomes}c, lower image).

\subsection{Design Goals}
The research methods contributed valuable insights into different aspects of the design space spanning visual interpretation of texts and the construction of scholarly arguments. The co-design workshop highlighted how visual aids can support and externalize interpretive reasoning during close reading. Our speculative sketches helped envision and discuss interface configurations and functionalities inspired by both workshop findings and prior research. Finally, rapid prototyping allowed us to test these interface ideas and investigate their potential for supporting hybrid forms of textual analysis. 

Drawing on these insights, we formulate three design goals (DG) to guide the development of the Textarium environment: 
\begin{enumerate}
    \item Support and augment the text interpretation process through visualizations and computational techniques that integrate close and distant reading practices.
    \item Employ lightweight and legible text analysis algorithms that enhance interpretive agency without introducing excessive complexity or opacity.
    \item Enable the integration of interpretive outcomes—annotations and abstractions—into self-authored, interactive scrollytellings to connect analysis with scholarly communication.
\end{enumerate}

\section{Textarium}

Arguments are an essential part of every scholarly discourse. In text-based digital humanities research, the formulation of an argument is often preceded by close and/or distant reading practices, including annotation, abstraction, and iterative interpretation. Interactive text visualizations can support this process by offering interfaces for exploring, annotating, and reorganizing textual material (DG1). However, scholarly publications still commonly rely on static representations of this process, limiting the communicative potential of dynamic interpretative acts. 

\begin{figure}
    \centering
    \includegraphics[width=\linewidth]{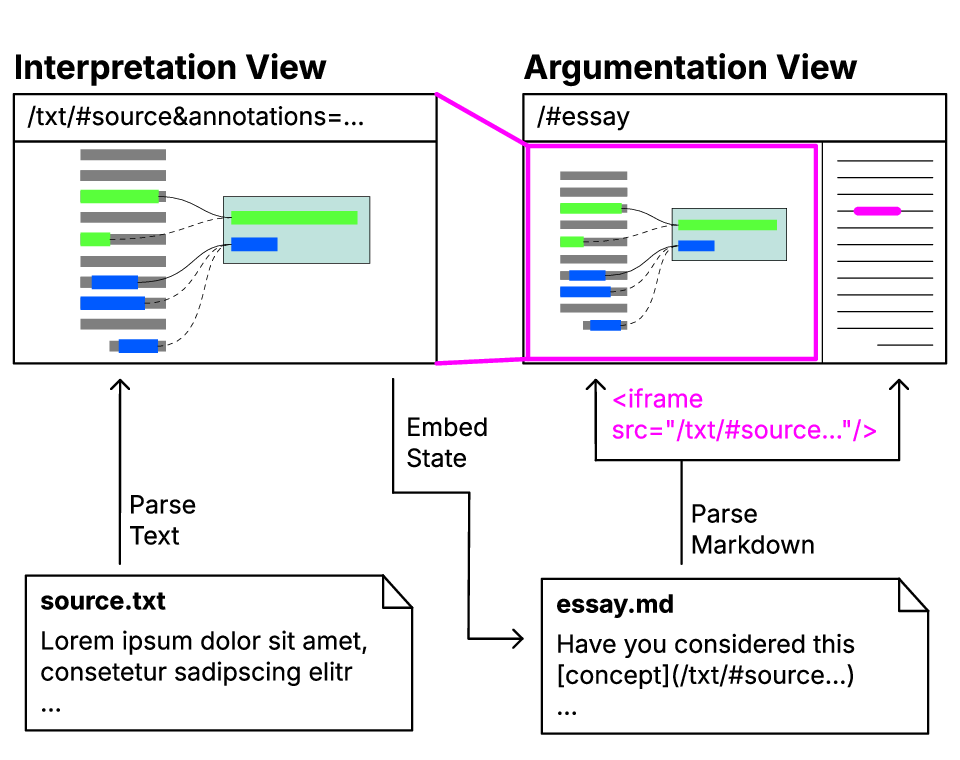}
    \vspace{-0.3in}
    \caption{Parameterized visualizations can be embedded as interactive elements within a narrative argumentation view.}
    \label{fig:embedding-logic}
\end{figure}

\textit{Textarium} was developed as a web-based environment\revised{\footnote{\url{https://uclab.fh-potsdam.de/textarium/}}} with the aim to bridge interpretation and communication: it enables scholars to annotate and abstract interpretive insights from a text and then embed these evolving visualizations into interactive, authored narratives (DG3). At the core of the system is a mechanism that parameterizes interactions during interpretation and encodes them in shareable URLs, which can be embedded directly into digital scholarly writing~\cite{eingartner_inflecting_2025}, making the visual reasoning process both accessible and reproducible (DG3) (see \autoref{fig:embedding-logic}).

\subsection{Overview}

Textarium consists of two interconnected views, each supporting a different phase of scholarly work while remaining tightly linked:

\begin{itemize}
    \item The \textbf{Interpretation View} supports annotation, comparison, and abstraction of textual elements. Readers can actively engage with the text by highlighting passages, establishing conceptual links, and exploring textual patterns through lightweight visual augmentations (DG1, DG2).
    \item The \textbf{Argumentation View} enables scholars to weave these interpretive insights into narrative form. Here, arguments are written and structured while embedding live, interactive references to the interpretations developed earlier (DG3).
\end{itemize}

This two-part structure was devised to support an entangled reading–writing process, allowing scholars to iteratively move between interpretative analysis and communicative synthesis. 
\revised{While conceptually intertwined, these activities operate on distinct texts: the original source and the emerging argument. The interface aims to support fluid transitions between both views, making the interpretative lineage of claims both visible and navigable.
} The design of the interface was informed by the aforementioned objectives emphasizing scholarly agency, conceptual clarity, and transparent integration of algorithmic support.

\subsection{Interactions}

The interaction design of Textarium is grounded in practices from literary and textual analysis, combining manual interpretation with computational augmentation (DG1). Interactions in the interpretation view are not merely functional—they externalize and structure interpretive acts—from annotation to abstraction—in ways that can be revisited, shared, and discussed. Likewise, the argumentation view is not intended to close off a discussion, but to open up a possibility space for multiple readings of a text (DG3).

\subsubsection{Interpretation}

\begin{figure}[!hbt]
    \centering
    \frame{\includegraphics[width=\linewidth]{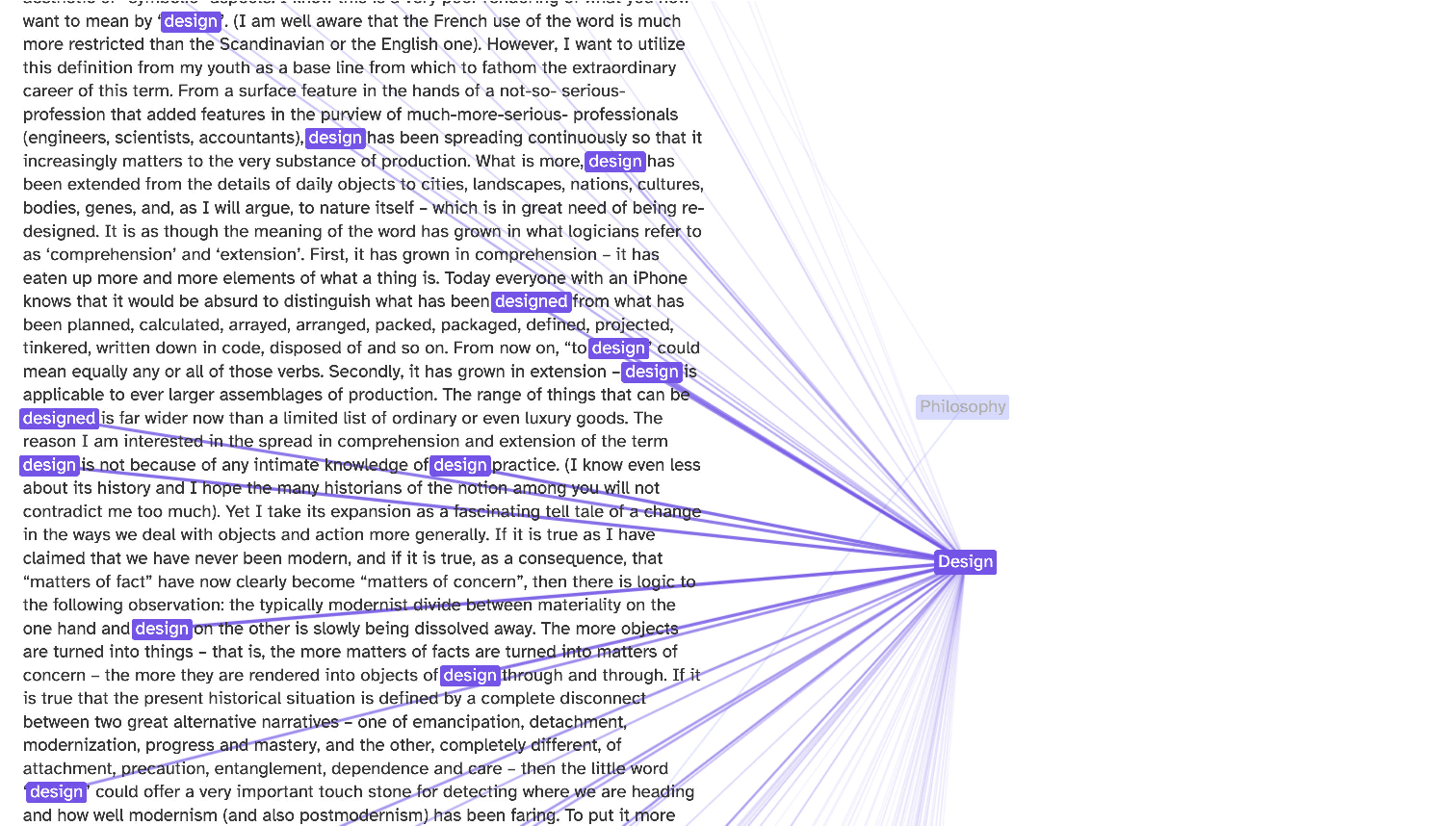}}
    \vspace{-0.25in}
    \caption{Annotated words are linked to extracted tokens and connected visually by dynamic lines. }
    \label{fig:annotation-example}
\end{figure}

Annotations serve as the initial interpretive layer. Scholars highlight individual words or short phrases within the source text. These selections are extracted into a separate pane for further examination and can be spatially rearranged (see \autoref{fig:annotation-example}). Visual connections between the original location in the text and the extracted annotations are rendered using lines, reinforcing the traceability of interpretive focus. Scrolling the source text dynamically adjusts the lines to subtly convey the distribution of selected terms across the entire text.
\revised{To ensure visual coherence, the highlights for different annotations, the respective extracts, on the right side, as well as the lines connecting them cycle through the colors of the IBM color-blind safe palette, designed for perceptual distinctiveness and accessibility.}
To extend the reader’s attention, 
\revised{the environment automatically highlights additional textual instances that share the same stem as a selected word (DG2)—e.g., highlighting “design” will also surface “designed” and “designing.” These highlights appear instantly during interpretation, supporting pattern recognition and interpretive exploration without interrupting the scholar’s reading flow.}
This functionality relies on lightweight and interpretable text analysis techniques—such as stemming, regular expression matching, and string similarity. These connections assist in identifying recurring motifs or shifts in rhetorical usage while maintaining transparency of method (DG2). Importantly, each annotation is encoded in human-readable form in the URL, allowing interpretations to be preserved, shared, and embedded (DG3).

\begin{figure}[!hbt]
    \centering
    \frame{\includegraphics[width=\linewidth]{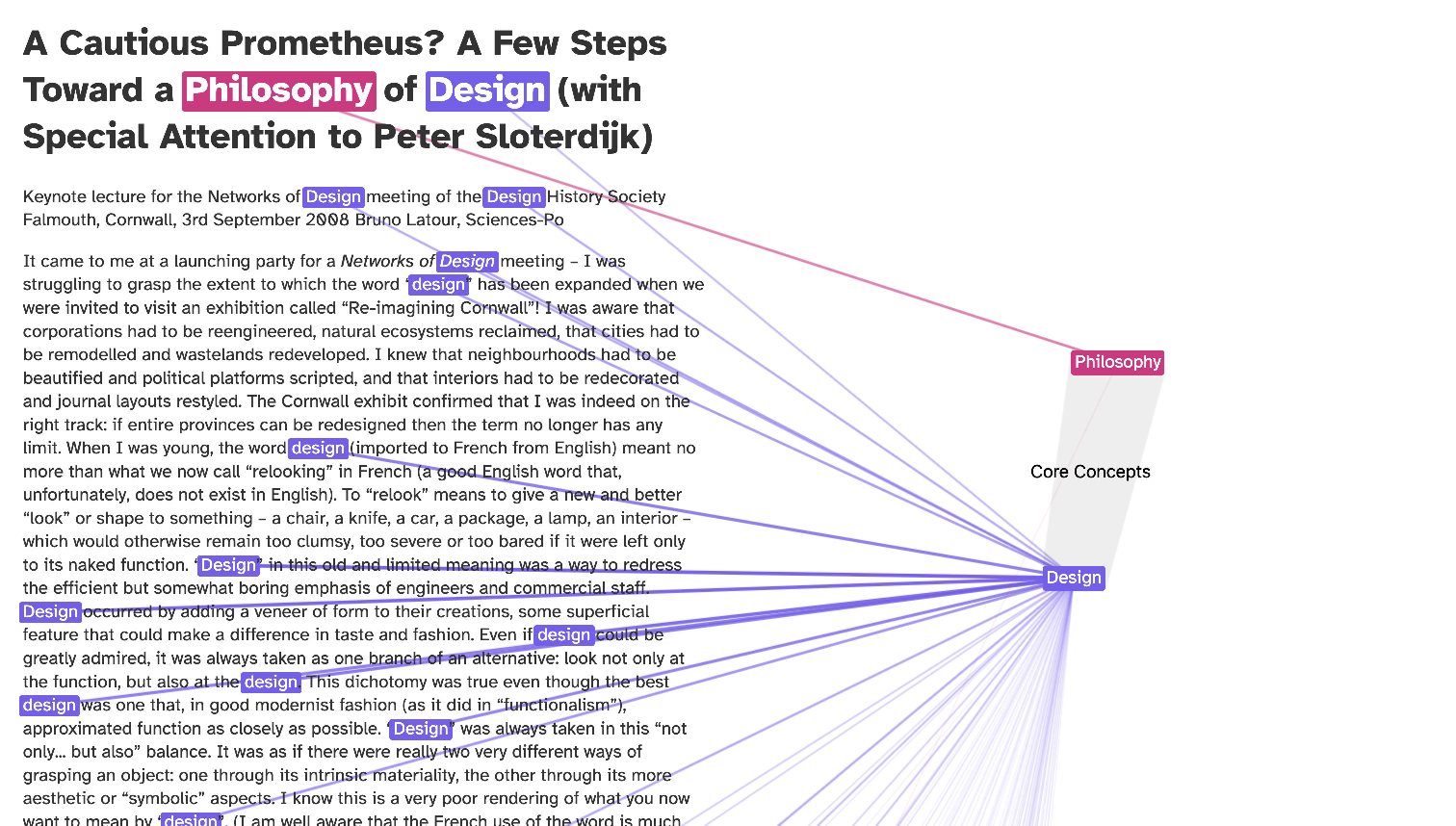}}
    \vspace{-0.25in}
    \caption{Annotations can be abstracted into conceptual groupings.}
    \label{fig:abstraction-example}
\end{figure}

Beyond individual annotations, readers may begin to detect conceptual groupings—collections of textual fragments that resonate semantically or rhetorically. Textarium supports the creation of such abstractions by allowing the selection and merging of annotations into higher-order constructs (see \autoref{fig:abstraction-example}) (DG1). These groupings can be named (e.g., “languages”) and are visually encoded in a \revised{light gray hull} to indicate conceptual cohesion.
The environment also provides suggestions for possible abstractions based on linguistic similarity (DG2). However, the final grouping remains in the scholar’s hands, reinforcing interpretative control and human-centered meaning-making (DG1). Like annotations, abstractions are encoded in the URL, preserving not only the analytical output but also the reasoning structure behind it (DG3).

\subsubsection{Argumentation}

\begin{figure}[!hbt]
    \centering
    \frame{\includegraphics[width=\linewidth]{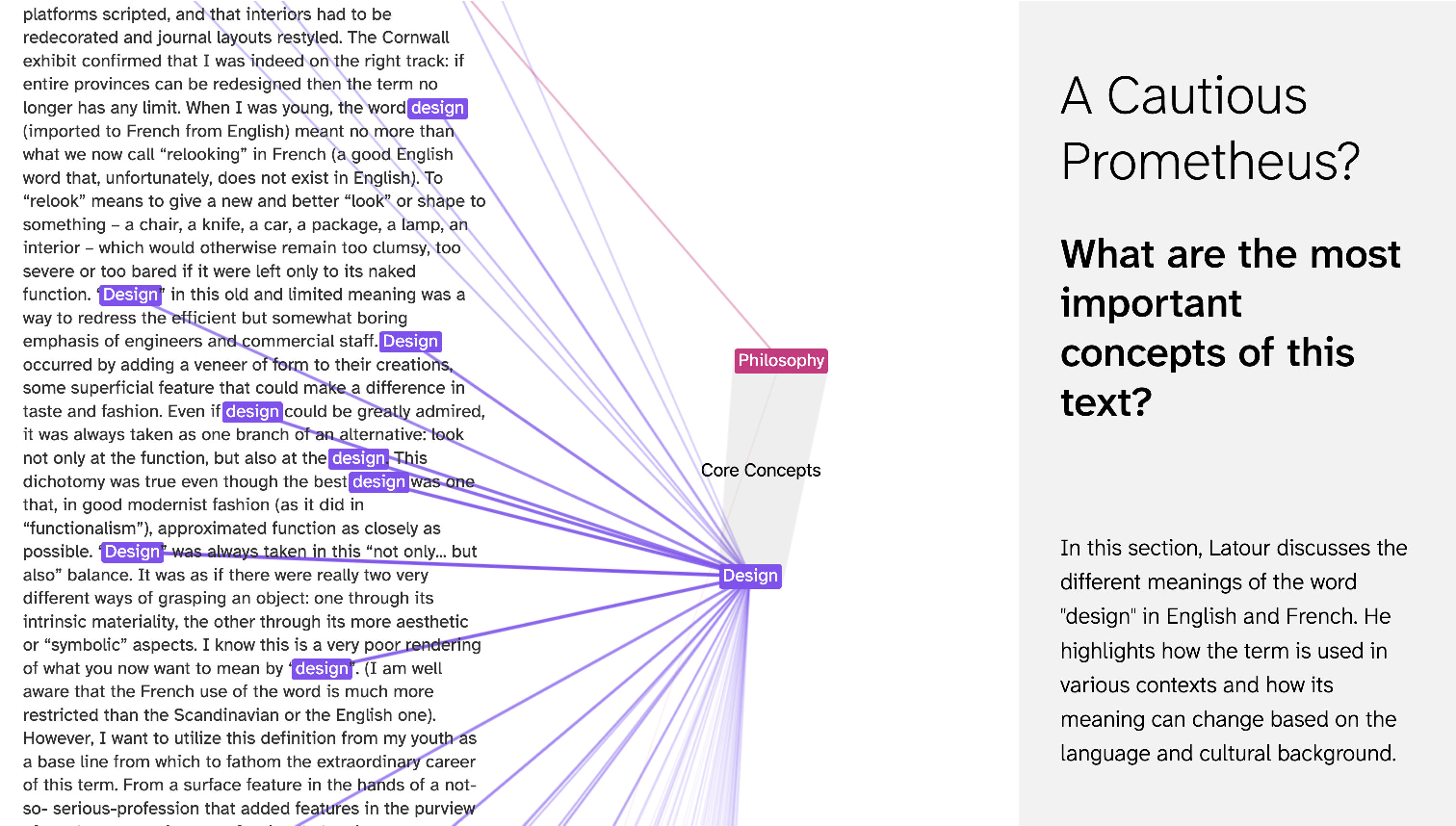}}
    \vspace{-0.25in}
    \caption{Specific annotations as well as interpretive abstractions can be embedded in a scholarly, scrollable essay.}
    \label{fig:argumentation-example}
\end{figure}

During argumentation, scholars compose narrative texts that contextualize and articulate the interpretive findings (see \autoref{fig:argumentation-example}). Each section of the argument can include live, embedded visualizations—linking back to specific interpretative states (DG3).
This view follows a scrollytelling format: as the reader moves through the argument, embedded frames dynamically load and display the relevant interpretive configuration. This enables readers not just to follow the scholar’s reasoning, but to explore and inspect the underlying interpretive choices. Through this integration, interpretation becomes part of the argument itself—not merely an illustration, but a structural element of scholarly communication (DG3).
By exposing the underlying interpretative traces in an interactive format, readers are invited to engage critically with the author’s decisions and explore alternative perspectives (DG1, DG3).

\subsection{Web-based Architecture}

To realize these interactions, Textarium builds on existing web technologies and the modular \textit{!nflect} framework \cite{eingartner_inflecting_2025}. The environment is organized into separate folders for each view: the interpretation view resides in the subdirectory \texttt{/txt} dedicated to engagements with textual sources, while the argumentation view is composed of Markdown-based content rendered by the \texttt{index.html} in the root folder \texttt{/} (DG3).
Interpretive actions—such as annotation and abstraction—are encoded in the URL hash, ensuring that no server-side storage is needed and that each interpretive state remains portable and transparent \cite{berners-lee_uniform_2005} (DG2, DG3). These URLs can be embedded into the narrative text as hyperlinks. The Markdown parser used in the argumentation view automatically detects such links and renders them in an \texttt{<iframe>} when the relevant section comes into view.
This architecture supports reproducibility, shareability, and modular composition. It enables scholars to embed their interpretations directly into digital narratives—without requiring technical expertise or infrastructure (DG1, DG3).

\section{Discussion}
In this paper we have introduced Textarium, a prototypical environment aimed at exploring new ways of connecting visual text interpretation with scholarly argumentation. Our goal was to experiment with how lightweight algorithmic augmentations and subtle visual encodings could support the processes of annotation and abstraction, while remaining accessible and legible to humanities scholars. By parameterizing these interpretive interactions and enabling their integration into narrative formats, Textarium shows how interactive visualizations can be more directly embedded in scholarly reasoning and communication. While still in an early stage, the system points toward possibilities for combining close and distant reading practices within an environment that foregrounds interpretive agency and scholarly transparency.

Although Textarium currently relies on rule-based techniques for textual augmentation, we recognize that more advanced machine learning models—particularly large language models (LLMs)---are increasingly shaping the landscape of digital text analysis. Rather than rejecting such tools, we see opportunities to explore how interactive and visual environments might make their influence more traceable and open to reflection. Parametrization, in this context, could help preserve interpretive paths, enable reproducibility, and foster a more critical engagement with computational outputs. If and how such systems might support dialogic or exploratory modes of scholarship remains an open question. We believe that combining visualization and narration offers a promising framework for surfacing ambiguity, negotiating meaning, and interrogating model behavior---especially in the context of complex or contested texts.

An important aspect of this work is the potential to reveal different interpretations and highlight interpretive differences. Rather than enforcing consensus, Textarium supports parallel and potentially conflicting perspectives that can coexist within the same analytical framework. The ability to parameterize and share specific interpretive states allows scholars to document their reasoning and respond to, reframe, or challenge the reasoning of others. Thus, disagreement becomes a productive force rather than an obstacle to be resolved. We see the potential for epistemic tension----including those between human interpretations and machine readings---as central to fostering more reflective and inclusive scholarly discourse. 

Future work will be needed to better understand the possibilities and limits of this approach. We plan to experiment with Textarium in a range of interpretive settings to see how different scholarly communities engage with the system. Additional interaction techniques could support more flexible ways of organizing, layering, or abstracting interpretive constructs. We are also curious about reintroducing more free-form or expressive forms of annotation, inspired by analog reading practices, that might allow for more fluid and flexible engagements with the text. Ultimately, this project remains exploratory. Rather than offering a finalized solution, we hope to provoke further questions about how interpretation, computation, and communication might be brought into closer dialogue in the digital humanities.

\bibliographystyle{abbrv-doi}

\bibliography{refs}
\end{document}